\documentclass[onecolumn,pra,superscriptaddress,10pt,letterpaper]{article}
\usepackage{amsmath,graphicx,amsfonts,bm,amssymb,mathrsfs}
\usepackage{xcolor}
\usepackage{times}
\usepackage{opex3,cite}
\begin{document}

\title{Cavity QED implementation of non-adiabatic holonomies for universal quantum gates in decoherence-free subspaces with nitrogen-vacancy centers}

\author{Jian Zhou,$^{1,2}$ Wei-Can Yu,$^{2}$ Yu-Mei Gao,$^{3}$ and Zheng-Yuan Xue$^{2,*}$}

\address{$^1$ Anhui Xinhua University, Hefei, 230088, China \\
$^2$ Guangdong Provincial Key Laboratory of Quantum Engineering and Quantum Materials, School of Physics and Telecommunication Engineering,  South China
Normal University, Guangzhou 510006, China\\
$^3$ Zhongshan College, University of Electronic Science and Technology of China, Zhongshan, 528402, China}

\email{*zyxue@scnu.edu.cn}

\date{\today}

\begin{abstract}
A cavity QED implementation of the non-adiabatic holonomic quantum computation in decoherence-free subspaces is proposed with nitrogen-vacancy centers coupled commonly to the whispering-gallery mode of a microsphere cavity, where a universal set of quantum gates can be realized on the qubits. In our implementation, with the assistant of the appropriate driving fields, the quantum evolution is insensitive to the cavity field state, which is only virtually excited. The implemented non-adiabatic holonomies,  utilizing optical transitions in the $\Lambda$ type of three-level configuration of the nitrogen-vacancy centers, can be used to construct a universal set of quantum gates on the encoded logical qubits. Therefore, our scheme opens up the possibility of realizing universal holonomic quantum computation with cavity assisted interaction on solid-state spins characterized by long coherence times.
\end{abstract}



\section{Introduction}

Quantum computation is an attractive field on precision control of quantum systems. But quantum systems are inevitably influenced by the decoherence effect induced from the environment, which stands in the way of physical implementation of quantum computation. In order to overcome this difficulty, there are many strategies to correct or avoid errors during the implementation of the gate operations. The geometric phases \cite{Berry1984, Aharonov1987} and their non-Abelian extensions, quantum holonomies \cite{Wilczek1984, Anandan1988}, accompanying in evolutions of quantum systems, unveil important geometric structures in the description of the dynamics of quantum states. Geometric phases  depend only on global geometric properties of the evolution paths so that they are largely insensitive to many types of local noise. Based on this distinct merit, holonomic quantum computation (HQC), first proposed by Zanardi and Rasetti \cite{Zanardi1999}, has emerged as a promising strategy to implement universal quantum computation in a robust way  \cite{Duan2001a, Zhu2003a, Zhang2006, Thomas2011}.

It is well known that the geometric phases, either Abelian or non-Abelian, consist of both adiabatic and non-adiabatic parts. In adiabatic evolution, it requires that quantum evolutions fulfill the famous adiabatic condition, and thus quantum gates constructed in this way generally have very slow speeds. This is unacceptable as the time needed for an adiabatic quantum gate maybe on the order of the coherence time of the used quantum two-level systems in typical systems \cite{Wang2001,Zhu2002}. Therefore, one needs to consult quantum gates based on non-adiabatic evolution \cite{Wang2001,Zhu2002}, which does not require the  adiabatic condition so that the gate speeds will only depend on the merits of the employed quantum systems. In this case, the built-in fault-tolerance of the non-adiabatic geometric phases provide a more practical way in implementing quantum computation. Therefore, many renewed efforts have recently been given in this direction both theoretically and experimentally \cite{Sjoqvist2012, Abdumalikov2013, Feng2013}.

Meanwhile, as an another promising way to avoid the effect of decoherence, decoherence-free subspace (DFS) can suppress the collective dephasing noise caused by the interaction between quantum systems and their environment \cite{Duan1997a, Zanardi1997, Lidar1998, Kielpinski2001, Ollerenshaw2003, Bourennane2004}. Therefore, many efforts have been paid to combine the HQC with DFS encoding \cite{Wu2005, Xu2012, Liang2014a, Zhang2014d, xu2014}. In this way, one can consolidate the best of the two quantum computation strategies, i.e.,
resilient against environment induced collective decoherence of the DFS approach and the operational robustness of the HQC against local noises.

Here, we propose a non-adiabatic HQC (NHQC) scheme with DFS encoding based on nitrogen-vacancy (NV) centers coupled commonly to the whispering-gallery mode (WGM)  of a fused-silica microsphere cavity. The NV center system is considered as a promising candidate for physical implementation of quantum computation, due to its sufficiently long electronic spin lifetime as well as the possibility of coherent manipulation even at room temperature \cite{Stoneham2009, shi2010}.  Since the electronic spins of the NV centers can be well initialized and manipulated in an optical way, quantum gates acting on the single-spin state can be obtained with very high efficiently \cite{Jelezko2004, Jiang2009}. Based on the symmetry structure of the system-environment interaction, we propose an encoding method of the logical qubits in the DFS, where any logical qubit state that undergoes a cyclic evolution will end up with a new state in the same subspace, without going out of the subspace. Moreover, based on the numerical simulation under realistic conditions, we show that our NHQC scheme in DFS can realize a universal set of quantum gates with high fidelity.

\begin{figure}[tbp]
\centering
\includegraphics[width= 9cm]{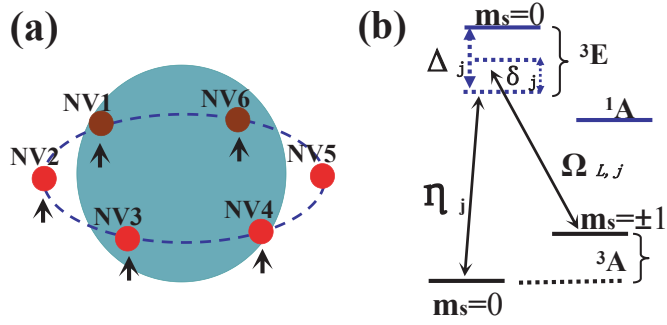}
\caption{(a) Schematic setup of the fused-silica microsphere cavity, where
$N$ identical NV centers (red dots) in diamond nanocrystals are equidistantly attached around the equator of the cavity. (b) Level diagram for a NV center, with $g$ ($\Omega _{L}$) is the coupling strength between NV center and the WGM (laser pulse). We encode a physical qubit in the subspace spanned by the states of $m_{s}=0$ and $m_{s}=-1$.} \label{setup}
\end{figure}

\section{Setup and effective Hamiltonian}

The setup we consider and the energy level configuration of the NV centers are schematically shown in Fig. \ref{setup}. In the cavity, the lowest-order WGM, corresponding to the light traveling around the equator of the microsphere, offers exceptional mode properties for reaching strong light-matter coupling. As shown in the Fig. \ref{setup}(a), $N$ NV centers, from separate diamond nano-crystals, are strongly coupled to the WGM of a microsphere cavity \cite{Park2006, Barclay2009}. The NV centers we considered can be modeled as three-level systems, as shown in Fig. 1(b), where the states $\left\vert ^{3}A,m_{s}=0\right\rangle $ and $\left\vert ^{3}A,m_{s}=-1\right\rangle $ are encoded as our qubit states $\left\vert 0\right\rangle $ and $\left\vert 1\right\rangle $, respectively. The state $\left\vert ^{3}E,m_{s}=0\right\rangle $ is labeled by $\left\vert e\right\rangle $ and leave aside the metastable $^{1}A$ state, which has not yet been fully understood \cite{Manson2006}. In our implementation, the optical transition $\left\vert 0\right\rangle \rightarrow \left\vert e\right\rangle$ and $\left\vert 1\right\rangle \rightarrow\left\vert e\right\rangle $ (with transition frequencies $\omega_{e0}$ and $\omega_{e1}$) are coupled by the WGM with frequency
$\omega _{c}$ and a classical laser filed with frequency $\omega _{L}$ \cite{Santori2006}, respectively. Both coupling is far-off resonant from their transition frequencies  so  that the $|e\rangle$ state can be adiabatically eliminated. Note that there is a degenerate state $\left\vert ^{3}A,m_{s}=1\right\rangle$ for the qubit state $\left\vert ^{3}A,m_{s}=-1\right\rangle$, these two degenerated states can be  selectively addressed by the polarization of the optical field \cite{Santori2006}.  The NV centers are fixed and the distance of two NV centers is much larger than the wavelength of the WGM, so that each driven laser field, with frequency $\omega_L$ and initial phase $\varphi_j$, can interact individually with an NV center and the direct coupling among the NV centers can be negligible. Then the interaction of the whole quantum system, in units of $\hbar =1$, can be written as
\begin{eqnarray}\label{hs}
H_0&=&\omega_c a^{\dag} a +  \sum_i \omega_i |i\rangle\langle i|,\notag\\
H_{S}&=&H_0 +\sum _{j=1}^{N}\left[\eta_{j} a|e\rangle_j\langle 0|+
\Omega_{L,j}e^{-i(\omega_{_L} t+\varphi_j)}|e\rangle_j\langle 1| + \text{H.c.}\right],
\end{eqnarray}
where $a^{\dag}(a)$ is the creation (annihilation) operator of the WGM of the cavity, $\omega_i$ is the frequency of $i$th energy level of the identical NV centers with $i\in\{0, 1, e\}$, $\eta_j$ ($\Omega_{L,j}$) is the coupling strength between $j$th NV center and the cavity (laser). In the interaction picture with respect to the free Hamiltonian $H_0$, under the rotating-wave approximation (RWA), the interaction  can be written as
\begin{equation} \label{int}
H_{I}=\sum _{j=1}^{N} g_{j}\left(a\sigma _{j}^{+}e^{-i (\delta_{\pm,j} t -\varphi_{j})}+\text{H.c}.\right),
\end{equation}
where we have assumed $\eta_j=G$ for simplicity, the effective cavity assisted interaction strength are $g _{j}=G \Omega _{L,j}(\frac{1}{\Delta+\delta _{\pm,j}}+\frac{1}{\Delta})$ with $\Delta=\omega _{e0}-\omega
_{c}$, $\sigma _{j}^{+}=\left\vert 1\right\rangle_{j} \left\langle 0 \right\vert $,
$\sigma _{j}^{-}=\left\vert 0\right\rangle_{j} \left\langle 1\right\vert $, and $\delta_{\pm,j}=\pm(\omega _{c}-\omega _{10}- \omega_{L,j})$.

When two laser fields are applied to a pair of NV centers (for example the $m$th and the $n$th), one can obtain the effective Hamiltonian, under the condition of $\delta_{\pm}\gg g_{mn}$, as
\begin{eqnarray}\label{eff1}
H_{m,n}=\sum \limits_{j=m,n} \frac{g^{2}_{mn}}{\delta_{\pm}} (-a^{\dag}a|0\rangle_{j}\langle 0|+aa^{\dag}|1\rangle_{j}\langle 1|)
+\frac{g^{2}_{mn}}{\delta_{\pm}}(e^{i\varphi_{mn}}\sigma_{m}\sigma^{+}_{n}+\mathrm{H.c.}),
\end{eqnarray}
where $\varphi_{mn}=\varphi_{m}-\varphi_{n}$. Neglecting the level shift terms, which can be compensated by using additional lasers \cite{Tamarat2006, Acosta2012}, the effective Hamiltonian between the two NV centers reduces to
\begin{eqnarray}\label{H}
H_{m,n}=\frac{g^{2}_{mn}}{\delta_{\pm}}(e^{i\varphi_{mn}}\sigma_{m}\sigma^{+}_{n}+\mathrm{H.c.}).
\end{eqnarray}
With this Hamiltonian, we next show how a universal set of non-adiabatic holonomic
gates can be implemented.

\section{Non-adiabatic holonomic One-qubit logical gates}

Now we turn to construct universal single logical qubit operations
for the NHQC in DFS. It is noted that a logical qubit consists of two
physical qubits is not sufficient for a dephasing environment \cite{Xu2012}. We here utilize three physical qubits to encode a logical qubit.  The interaction between three physical qubits (NV centers) and the dephasing environment can be described by the interaction Hamiltonian, $H_{I}=S^{z}\bigotimes{E}$, where
$S^{z}=\sum^{3}_{i=1}\sigma_{i}^{z}$ is the collective dephasing operator and $E$ is an arbitrary environment operator. In this case, there exist a three-dimensional DFS of $$S_1=\{|100\rangle,|001\rangle,|010\rangle\},$$
where the computational basis, i.e., the logical qubit states, are
encoded as $|0\rangle_{L}=|100\rangle$, $|1\rangle_{L}=|001\rangle$, and  $|a_{1}\rangle=|010\rangle$ is used as a third ancillary state.

In order to implement the single logical qubit operation, we apply operation $U_{12}(t)=\exp[-i \int_0^{\tau_1} H_{12} dt]$, which can be obtained from Eq. (\ref{H}) for an operate time of $\tau_{1}$, to the NV centers $1$ and $2$ by tuning laser beams with the same detuning $\delta_+$ from their respectively transition and phase difference $\varphi=\varphi_{1}-\varphi_2$. Meanwhile, we also apply $U_{23}(\tau_1)=\exp[-i \int_0^{\tau_1} H_{23} dt]$ on NV centers $2$ and $3$ by tuning laser beams with the same detuning $\delta_-$ and identical phase.
Then, in the DFS $S_1$ the effective Hamiltonian reads
\begin{eqnarray}\label{h1}
H_{1}=\lambda_{1}\left(\sin\frac{\theta}{2}e^{i\varphi}|a_{1}\rangle_{L}\langle0|
-\cos\frac{\theta}{2}|a_{1}\rangle_{L}\langle1|\right) +\mathrm{H.c.},
\end{eqnarray}
where the effective Rabi frequency
$\lambda_{1}=\sqrt{|g_{12}|^{4}+|g_{23}|^{4}}/\delta$ with $\delta=|\delta_{\pm}|$
and the phase $\theta=2 \arctan\left(|g_{12}|^{2}/|g_{23}|^{2}\right)$ can be tuned by the amplitude of the incident lasers. This Hamiltonian, consisting of a WGM and three physical qubits (three NV centers), is in the $\Lambda$ type with ancillary logical state $|a_{1}\rangle_{L}$ at the top while the logical qubit states $|0\rangle_{L}$ and $|1\rangle_{L}$ at the bottom. In the dressed state representation, the two degenerate states of Eq. (\ref{h1}) are
\begin{eqnarray}
|d\rangle_{L}&=&\cos\frac{\theta}{2}|0\rangle_{L}+\sin\frac{\theta}{2}e^{i\varphi}|1\rangle_{L}, \notag\\
|b\rangle_{L}&=&\sin\frac{\theta}{2}e^{-i\varphi}|0\rangle_{L}-\cos\frac{\theta}{2}|1\rangle_{L}.
\end{eqnarray}
It is obviously that the dark state $|d\rangle_{L}$ decouples from the 'bright' state $|b\rangle_{L}$ and the excited state $|a_{1}\rangle_{L}$, while the
'bright' state $|b\rangle_{L}$ couples to the excited state
$|a_{1}\rangle_{L}$ with the effective Rabi frequency $\lambda_{1}$.

As a result, the evolution operator $U_1(\tau_1)=\exp(-i\int_0^{\tau_1} H_1 dt)$ can realize the holonomic gates with following two conditions. The first one is that controlling the operation time to meet $\lambda_1 \tau_1=\pi$. This condition will ensure the states evolving in the subspace $|\psi_i(t)\rangle_L=U_1(t)|i\rangle_L (i=d,b)$ undergo the cyclic evolution $|\psi_i(\tau_1)\rangle_L=|\psi_i(0)\rangle_L$. Furthermore, the relation $\langle\psi_i(t)|H_1|\psi_j(t)\rangle=0\texttt{ }(i,j={d,b})$ is always hold, which means the evolution is pure geometric and the parallel-transport condition is natural satisfied. Therefore, we can obtain the non-adiabatic holonomic single qubit gates in the space spanned by \{$|0\rangle_{L}$, $|1\rangle_{L}$\} as
\begin{equation}\label{u1}
U_{1}(\theta,\varphi)=\left(\begin{array}{ccc}
\cos{\theta}&\sin{\theta}e^{-i\varphi}\\
\sin{\theta}e^{i\varphi}&-\cos{\theta}
\end{array}\right),
\end{equation}
where $\theta$ and $\varphi$ can be chosen by tuning the amplitude and relative phase of laser beams, and thus a set of universal single qubit gates can be realized non-adiabatically. For examples, we can implement a Hadamard gate with $U_1(\pi/4,0)$ and a phase gate as $U_1(\pi/2, \pi/4)U_1(\pi/2,0)$.

\begin{figure}[tbp]
\centering
\includegraphics[width=13cm]{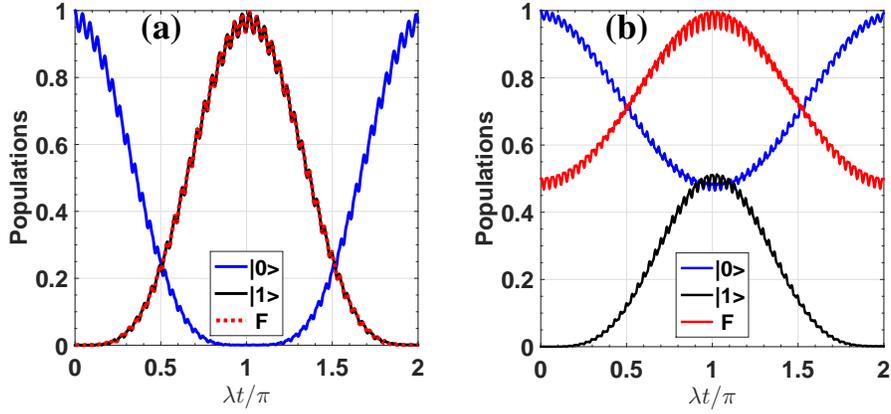}
\caption{ Qubit states population and fidelity under single-qubit operator $U_1(\theta, \varphi)$ with initial state $|0\rangle_L$, the red (dot) line represents the fidelity. (a) $\theta=\pi/2$ and $\varphi=0$. (b) $\theta=\pi/4$ and $\varphi=0$.}\label{single}
\end{figure}

\begin{figure}[tbp]
\centering
\includegraphics[width=12cm]{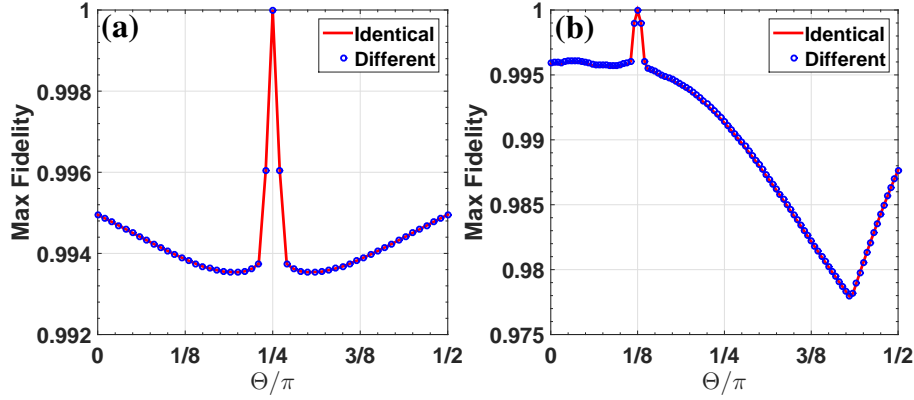}
\caption{ Maximum fidelity as the function of $\Theta$ in unit of $\pi$ with line (circle) the identical (different) decoherence of qubits under gate operations (a) $U_1(\pi/2, 0)$, (b) $U_1(\pi/4, 0)$.}\label{max1}
\end{figure}

Inevitably, the implementation process will suffer from  decoherence. Considering the main decoherence effect, the collective relaxation rate $\gamma$, dephasing rate $\gamma_{\phi}$ of NV centers and the decay rate $\kappa$ of the cavity, we simulate the performance of our scheme under realistic conditions with the Lindblad master equation \cite{Lindblad1976}
\begin{eqnarray}\label{master1}
\dot\rho=-i[H_I^{(3)}, \rho]+\frac 1 2 [\kappa\mathscr{L}(a)+\gamma\mathscr{L}(S^-)+\gamma_{\phi}\mathscr{L}(S^z)],
\end{eqnarray}
where $H_I^{(3)}$ is the Hamiltonian in the form of Eq. (\ref{int}) for the case of $N=3$ with the compensation of the Stark shift, $\rho$ is the density operator, $S^-=\sum_{i=1}^3 \sigma^-_i$ and $\mathscr{L}(\mathcal{A})=2\mathcal{A}\rho \mathcal{A}^\dagger-\mathcal{A}^\dagger \mathcal{A} \rho -\rho \mathcal{A}^\dagger \mathcal{A}$ is the Lindblad operator. In our simulation, we have used the following conservative set of experimental parameters.  The NV centers are located near the microcavity surface and the maximum coupling between an NV center and the cavity could be $G=2\pi\times 1$ GHz with the mode volume of $V_m=100\mu m^3$ \cite{Neumann2009}. For $\delta \ll \Delta$,  $g\simeq 2G \Omega_L/\Delta=2\pi \times 50$ MHz with $\Omega_L=2\pi \times 500$ MHz, $\Delta=2\pi \times 8$ GHz and $\delta=2\pi \times 1$ GHz to fulfill the condition of $\delta \gg g$. The qubit relaxation and dephasing rates are estimated to be $\gamma=\gamma_\phi \approx 2\pi \times 4$kHz \cite{Clark2003}.  The cavity decay rate is $\kappa=\omega_c/Q\simeq 2\pi \times 0.5$MHz with $Q=10^9$ \cite{Vernooy1998}. Assume that the logical qubit is initially prepared in $|0\rangle_L$ state while the cavity is in the vacuum state, the time-depend state populations and fidelity under the $X$ and Hadamard gates are depicted in Fig. \ref{single}(a) and \ref{single}(b) with the fidelity to be $99.5$ and $99.6$, respectively.  Note that, in the simulation, we use the Hamiltonian in Eq. (\ref{int}), and thus the obtained high fidelity also verifies the validity of the effective Hamiltonian in Eqs. (\ref{eff1}) and (\ref{H}). In addition, we assuming the collective dephasing of all the qubits, the violation of which will introduce very small infidelity to the gate operations, as shown in Fig. \ref{max1}(a) and \ref{max1}(b) for the two exemplified  gates $U_1(\pi/2,0)$ and $U_1(\pi/4,0)$ respectively, where we have investigated different cases by choosing different $\Theta$ in the initial state of $|\psi\rangle=\cos\Theta|0\rangle_L+\sin\Theta|1\rangle_L$. In the simulation, we have choose different decoherence of the physical qubits as $\gamma_1=\gamma_{\phi1}=0.8\gamma$, $\gamma_2=\gamma_{\phi2}=\gamma$ and $\gamma_3=\gamma_{\phi3}=1.2\gamma$, respectively.

\section{Non-adiabatic holonomic two-qubit logical gates}

The implementing of the holonomic one-qubit logical gates can be scalable to two-qubit scenario straightforwardly. For two logical qubits interacting collectively with the dephasing environment, there will exist a six-dimensional DFS $$S_2= \{|100100\rangle,|100001\rangle,|001100\rangle,
|001001\rangle,|101000\rangle,|000101\rangle\},$$
where the former and latter three physical qubits encode the first and second logical qubits, respectively. Then we can encode the logical qubit states same as that of the single qubit case, \emph{i.e.},
$|00\rangle_{L}=|100100\rangle$, $|01\rangle_{L}=|100001\rangle$, $|10\rangle_{L}=|001100\rangle$, and $|11\rangle_{L}=|001001\rangle$. Meanwhile, $|a_{2}\rangle=|101000\rangle$ and $|a_{3}\rangle=|000101\rangle$ are used as ancillary states.

\begin{figure}[tbp]
\centering
\includegraphics[width=13cm]{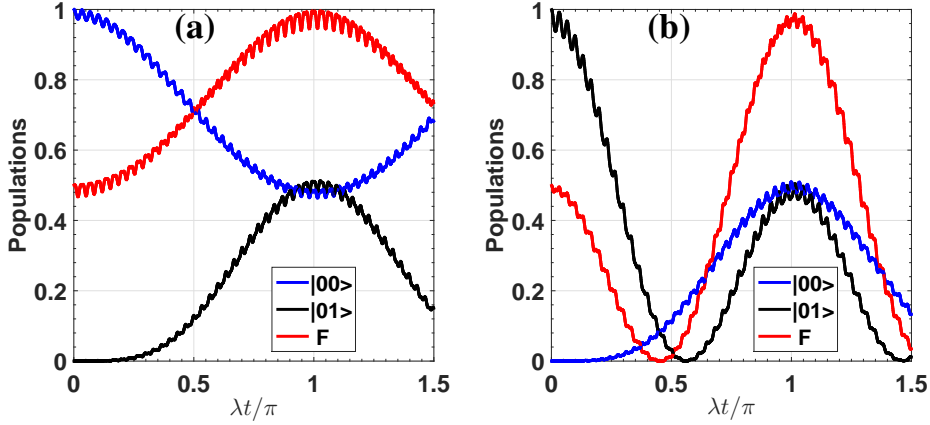}
\caption{ Qubit states population and fidelity under two-qubit operator $U_2(\pi/4, 0)$, the red line represents the fidelity. (a) Initial state of qubit $|00\rangle_L$. (b) Initial state of qubit $|01\rangle_L$.}\label{two}
\end{figure}

In order to implement the logical two-qubit operations, we apply the interaction in Eq. (\ref{H}) on the NV centers $3$ and $4$, by tuning laser beams with detuning $\delta_+$ and phase difference $\phi=\varphi_{3}-\varphi_4$, for the operation time  of $\tau_{2}$, i.e., $U_{34}(\tau_{2})=\exp[-i \int^{\tau_{2}}_{0} H_{34} dt]$. Meanwhile, we also apply $U_{36}(\tau_{2})=\exp[-i \int^{\tau_{2}}_{0} H_{36} dt]$ on the NV centers $3$ and $6$ by tuning laser beams with detuning $\delta_-$ and same phase. The total effective Hamiltonian then reads
\begin{eqnarray}
H_{2}=\lambda_{2}\left[\sin\frac{\vartheta}{2}e^{i\phi}(|a_{2}\rangle_{L}\langle00|
+|a_{3}\rangle_{L}\langle11|)
-\cos\frac{\vartheta}{2}(|a_{2}\rangle_{L}\langle01|
+|a_{3}\rangle_{L}\langle10|)\right] +\mathrm{H.c.},
\end{eqnarray}
where the effective Rabi frequency  $\lambda_{2}=\sqrt{|g_{34}|^{4}+|g_{36}|^{4}}/\delta$ with $\delta=|\delta_{\pm}|$
and the phase $\vartheta=2 \arctan\left(|g_{34}|^{2}/|g_{36}|^{2}\right)$ can be tuned by the amplitude of the incident laser. Obviously, the effective Hamiltonian can be decomposed to two commuting parts as $H_{2}=\lambda_{2}(H_a+H_b)$ with
\begin{eqnarray}
H_a=\sin\frac{\vartheta}{2}e^{i\phi}|a_{2}\rangle_{L}\langle00|
-\cos\frac{\vartheta}{2}|a_{2}\rangle_{L}\langle01|+\mathrm{H.c.}, \nonumber \\
H_b=\sin\frac{\vartheta}{2}e^{i\phi}|a_{3}\rangle_{L}\langle11|
-\cos\frac{\vartheta}{2}|a_{3}\rangle_{L}\langle10|+\mathrm{H.c.}.
\end{eqnarray}
Therefore, $\exp\left[-i\int^{\tau_{2}}_{0}H_{2}dt\right]
=\exp\left[-i\pi H_a\right]\exp\left[-i\pi H_b\right]$
under the $\pi$ pulse criterion $\lambda_{2}\tau_{2}=\pi$, which acts nontrivially on the computational subspace $\{|00\rangle_{L},|01\rangle_{L}\}$ and $\{|10\rangle_{L},|11\rangle_{L}\}$, respectively. That is, $H_{2}$, consisting of a WGM and six NV centers, effectively reduces to two $\Lambda$-like Hamiltonian with $H_a$ ($H_b$) acts on the top state $|a_{2}\rangle_{L}$ ($|a_{3}\rangle_{L}$) and  the bottom states $|00\rangle_{L}$ and$|01\rangle_{L}$ ($|10\rangle_{L}$ and $|11\rangle_{L}$). Analogous to the single-qubit gate case, the holonomic two-qubit logical gate in the subspace $\{|00\rangle_{L},|01\rangle_{L},|10\rangle_{L},|11\rangle_{L}\}$ can be written as
\begin{equation}\label{u2}
U_{2}(\vartheta,\phi)=\left(\begin{array}{cccc}
\cos{\vartheta}&\sin{\vartheta}e^{-i\phi}&0&0\\
\sin{\vartheta}e^{i\phi}&-\cos{\vartheta}&0&0\\
0&0&-\cos{\vartheta}&\sin{\vartheta}e^{-i\phi}\\
0&0&\sin{\vartheta}e^{i\phi}&\cos{\vartheta}
\end{array}\right).
\end{equation}

In general, we can realize nontrivial two-qubit holonomic logical gate in DFS by controlling the $\vartheta$ and $\phi$ separately, that is, adjusting the amplitude and phase of two laser beams, respectively. For example, acting $U_{2}(\vartheta,\phi)$ on logical qubit 1 and 2 and then acting $U_{1}(\theta,\varphi)$ on logical qubit $2$ with $\vartheta=\theta=\pi/4$ and $\phi=\varphi=\pi/2$, we can realize a CNOT gate in view of Eqs. (\ref{u1}) and (\ref{u2}). In addition, choosing the experimentally achievable parameters similar as in the single qubit case, numerical simulation of the populations and the fidelity of $U_2(\pi/4, 0)$ operation are shown in Fig. \ref{two}(a) and \ref{two}(b) with different initial state $|00\rangle_L$ and $|01\rangle_L$, respectively. The fidelity can reach about $99.5$ and $98.7$.

\section{Conclusion}
In summary, we have put forward a universal set of non-adiabatic holonomic gates in DFS by using NV centers coupled to the WGM of a cavity. By controlling the amplitude and relative phase of the driving lasers, we can realize arbitrary single-qubit and two-qubit operations. Numerical simulation shows that our scheme is stable to deviation of corresponding experimental parameters and insensitive to decoherence such as collective noises and local noises. Ultrahigh quality factor of the cavity and the exceptional spin properties of the NV centers make our scheme a promising candidate in experimental implementation of NHQC in DFS with high gate fidelity and short operation time.

\bigskip

\section*{Acknowledgments}
This work was supported by the NFRPC (No. 2013CB921804), the PCSIRT  (No. IRT1243), and the project of Anhui Xinhua university (No. 2014zr010).

\end{document}